# SPHERE on-sky performance compared with budget predictions

Kjetil Dohlen[*a], Arthur Vigan[a], David Mouillet[b], Francois Wildi[c], Jean-Francois Sauvage[a,d], Thierry Fusco[a,d], Jean-Luc Beuzit[b], Pascal Puget[b], David Le Mignant[a], Ronald Roelfsema[e], Johan Pragt[e], Hands Martin Schmid[f], Raffaele Gratton[g], Dino Mesa[g], Riccardo Claudi[g], Maud Langlois[h], Anne Costille[a], Emmanuel Hugot[a], Jared O'Neil[i], Juan Carlos Guerra[i], Mamado N'Diaye[j], Julien Girard[i], Dimitri Mawet[k], Gerard Zins[i]

[a] Aix Marseille Université, CNRS, LAM (Laboratoire d'Astrophysique de Marseille) UMR 7326, 13388 Marseille, France
[b] IPAG (UMR 5274) BP 53. F-38041 Grenoble, France
[c] Observatoire de Genève, CH-1290 Sauverny, Switzerland
[d] ONERA, 29 avenue de la Division Leclerc, 92322 Châtillon, France
[e] ASTRON, P.O. Box 2, NL-7990 AA Dwingeloo, The Netherlands
[f] Institute of Astronomy, ETH Zurich, CH-8092 Zurich, Switzerland
[g] INAF Osservatorio Astronomico di Padova, Vicolo dell'Osservatorio 5, I-35122 Padova, Italy
[h] CNRS/CRAL/Observatoire de Lyon/Université de Lyon 1, Lyon, France
[i] European Southern Observatory, Alonso de Cordova 3107, Vitacura, Santiago, Chile
[j] Space Telescope Science Institute, 3700 San Martin Drive, Baltimore, MD 21218, USA
[k] California Institute of Technology, 1200 E. California Blvd., Pasadena, CA 91125, USA

**ABSTRACT**

The SPHERE (spectro-photometric exoplanet research) extreme-AO planet hunter saw first light at the VLT observatory on Mount Paranal in May 2014 after ten years of development. Great efforts were put into modelling its performance, particularly in terms of achievable contrast, and to budgeting instrumental features such as wave front errors and optical transmission to each of the instrument's three focal planes, the near infrared dual imaging camera IRDIS, the near infrared integral field spectrograph IFS and the visible polarimetric camera ZIMPOL. In this paper we aim at comparing predicted performance with measured performance. In addition to comparing on-sky contrast curves and calibrated transmission measurements, we also compare the PSD-based wave front error budget with in-situ wave front maps obtained thanks to a Zernike phase mask, ZELDA, implemented in the infrared coronagraph wheel. One of the most critical elements of the SPHERE system is its high-order deformable mirror, a prototype 40x40 actuator piezo stack design developed in parallel with the instrument itself. The development was a success, as witnessed by the instrument performance, in spite of some bad surprises discovered on the way. The devastating effects of operating without taking properly into account the loss of several actuators and the thermally and temporally induced variations in the DM shape will be analysed, and the actions taken to mitigate these defects through the introduction of specially designed Lyot stops and activation of one of the mirrors in the optical train will be described.

## 1. INTRODUCTION

Since the first discovery of an extrasolar planet orbiting another star than ours some 20 years ago, exo-planetology has become a major field of astrophysical research. It is also one of the most challenging ones instrumentally. Among the various observational techniques used in this field, that of direct imaging is particularly challenging, calling upon extremely precise and high-order adaptive optics for real-time correction of atmospheric turbulence and instrumental aberrations combined with state-of-the art coronagraphic equipment for diffraction control.

---

[*] E-mail: kjetil.dohlen@lam.fr

The VLT-SPHERE (spectro-polarimetric high-contrast exoplanet research) [1] represents the first generation of dedicated direct imaging planet hunters. It was installed at Paranal and saw first light in May 2014 after ten years of development by a large consortium of European laboratories. The instrument went through extensive validation and testing over the six first months of its on-sky existence and was allowed to start its scientific life by the end of 2014. The first 18 months of science activity, over which some 40 peer-reviewed articles have been published, have proven the high quality of the instrument. Among the most extraordinary results one can mention the unprecedented sharpness of the AU Microscopii disk whose prominent features, also recognizable in HST images, are found to move outwards at breath-taking speeds [2].

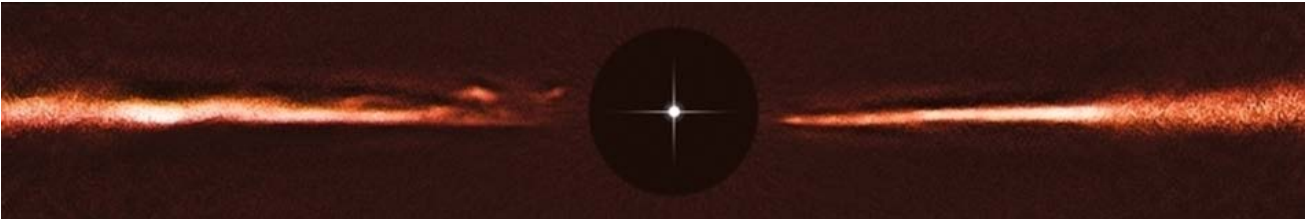

Figure 1. Unprecedented sharpness of disk images, such as for the AU Microscopii disk whose prominent features, also recognizeable in HST images, are found to move outwards at breath taking speeds [2].

The organs of the instrument are laid out on a large optical table as shown in Figure 2 (left). After passing through a fore-optics system containing a pupil tip-tilt mirror (PTTM), a derotator in the form of "K mirror" and polarization compensation optics, the beam enters a double relay system consisting of three stress-polished toroidal mirrors (TM1, TM2, TM3) [3] offering two pupil planes, for the fast image tip-tilt mirror (ITTM) and the deformable mirror (DM), respectively. The beam is then chromatically split by a dichroic plate reflecting the visible beam towards the wave front sensor (WFS) and the visible camera (ZIMPOL) [4] and transmitting the infrared light towards the dual imaging camera and spectrograph IRDIS [5] and the integral field spectrograph IFS [6]. The active components PTTM, ITTM, DM, and WFS compose the SAXO extreme AO system [7], see Figure 3. The overall shape of the bench is easily recognizable in the birds-eye view of the instrument in Figure 2 (right) as it sits on the Nasmyth platform of VLT UT3.

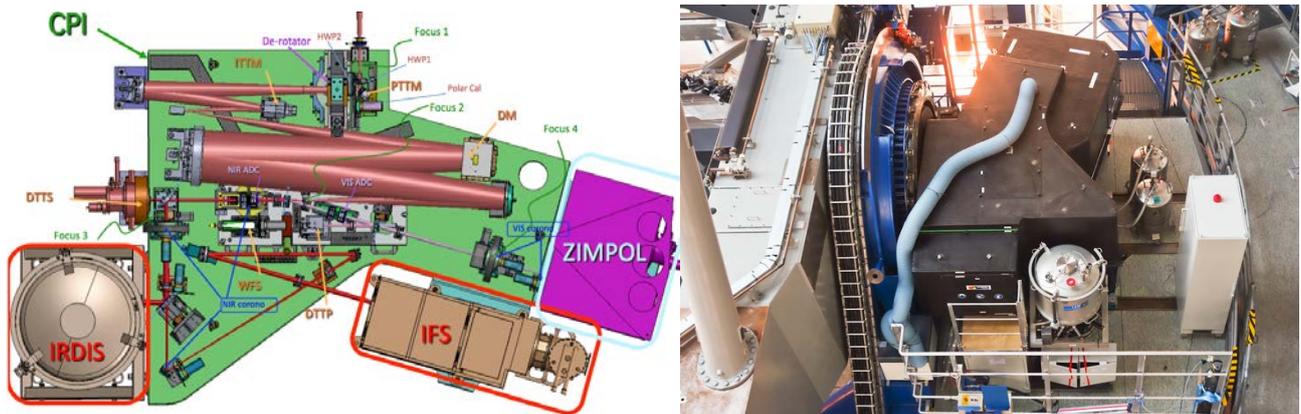

Figure 2. Conceptual design of SPHERE, left, showing the light paths through its various organs and into the three science channels IRDIS, IFS and ZIMPOL. Although covered by its dust, wind, and light-tight enclosure, its shape is easily recognizable in the bird's-eye view of the instrument on the VLT UT3 Nasmyth platform, right.

In this paper we describe the budgeting efforts related to instrumental features such as wave front errors and optical transmission, with the aim of comparing predicted with measured performance. In addition to comparing on-sky contrast curves and calibrated transmission measurements, we also compare the PSD-based wave front error budget with in-situ wave front maps obtained thanks to a Zernike phase mask, ZELDA, implemented in the infrared coronagraph wheel. We also discuss one of the most critical elements of the SPHERE system; its high-order deformable mirror, a prototype 40x40 actuator piezo stack design developed in parallel with the instrument itself. The development was a success, as witnessed by the instrument performance, in spite of some bad surprises discovered on the way. The devastating effects of operating without taking properly into account the loss of several actuators and the thermally and temporally induced variations in the DM shape will be analysed, and the actions taken to mitigate these defects through the introduction of specially designed Lyot stops and activation of one of the mirrors in the optical train will be described.

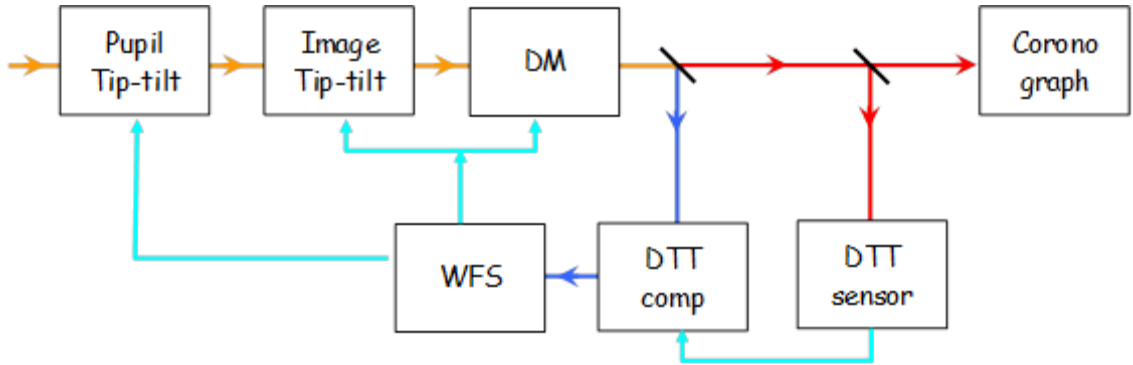

Figure 3. Block diagram of the SAXO extreme AO system of SPHERE illustrating its closed loop structure. The orange line represents light path for the full visible-to-near infrared bandwidth of the instrument, blue and red lines represent visible and infrared beams, respectively, and light blue lines indicate signal paths. While the ITTM and DM work at a frequency exceeding 1kHz, the DTT and PTT loops operate at much slower rates of 0.1 to 1 Hz.

## 2. WAVE FRONT ERRORS AND CONTRAST PERFORMANCE

While wave front budgeting is commonly done for most astronomical instruments, it is particularly important for high contrast imaging (HCI) instruments. But while classical instruments mainly care about the amount of light captured within the central area of the point spread function (PSF), the principal quality criterion for HCI concerns the distribution of the light scattered outside of the central core. In the low-aberration approximation using a perfect coronagraph, the distribution of light around the stellar core is proportional to the power spectral density (PSD) of the wave front. To fully explore the classical wave front error budget based on quadratic summing of a single number, the root-mean-square (RMS) error should therefor ideally be replaced by a 2-D budget. In practice, however, it is usually valid to assume that optical surface errors are isotropic, allowing us to use 1-D budgets.

Further simplifications are allowed by assuming a certain shape of the PSD for optical surfaces. Church [8] defines optical surface quality in terms of fractal finish, defining three different classes according to their spatial frequency power law: extreme ($f^{-2}$), Brownian ($f^{-3}$), and marginal ($f^{-4}$) fractal. In the search for typical PSDs valid for high-quality optical surfaces, we found that the current state of the art of optical polish, developed for UV lithography of electronic integrated circuits and reaching surface figure errors of 0.2nm RMS [9], generates extreme fractal surfaces. The cost of such polish quality applied to our prototype optics is exceeding, however, but we found that optical surfaces with 1 – 10nm RMS wave front error obtained using classical large-tools polishing techniques remain extreme fractal. For SPHERE we have therefore invested in high-quality classically polished mirrors, achieving down to 1-2nm RMS surface figure for which typical PSDs that we have measured do indeed follow the $f^{-2}$ law indicating extreme fractal surfaces, see Figure 7. Based on these results, we take the assumption of extreme fractal for all our surfaces. This leads to a further simplification of the budgeting work, since the RMS WFE of an extreme fractal surface is independent of the beam diameter on that surface [10].

The SPHERE wave front budget is illustrated in Figure 4. Three different spatial frequency domains can be recognized, corresponding to low (<4c/pup), mid, and high (>20c/pup). The ad-hoc spatial frequency unit of cycles per pupil diameter (c/pup) refers to aberrations in the form of sinusoidal wave front distortion with the given number of cycles across the pupil as projected onto a given surface. By Fourier analysis, the spatial frequency domain for wave front aberrations corresponds to the image plane in the imaging system, and an aberration of frequency $n$ c/pup gives rise to a speckle at a distance of $n$ $\lambda$/D from the star. The low frequency domain, up to 4 c/pup, corresponds to aberrations that were considered to be calibrated by phase diversity, while the mid frequency domain, covering up to 20 c/pup is corrected by the adaptive optics system. Residual aberrations in these domains correspond to un-calibrated non-common path aberrations, whose level is minimized by strict manufacturing tolerances on critical surfaces.

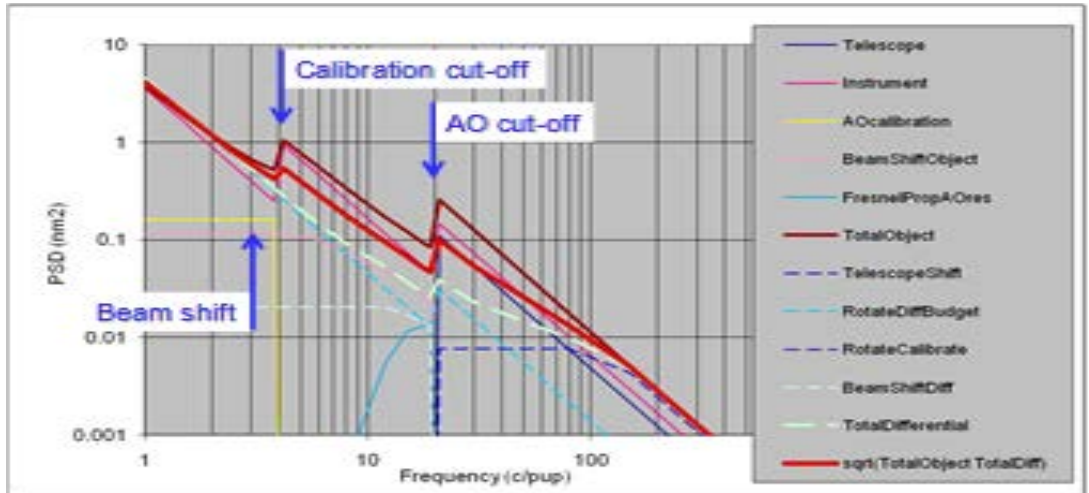
Figure 4. Illustration of the SPHERE wave front error budget, showing the PSD of individual surfaces as filtered by the adaptive optics system.

In addition to calibration errors, on-sky performance was expected to be mainly limited by chromatic beam shift on internal surfaces upstream of the dichroic separation between science and WFS beams, see Figure 5 [10], as well as the aberrations of the ADCs which are in the differential path and rotating during observation and to a lesser extent Fresnel propagation. These effects were expected to produce speckle de correlation at speeds varying with telescope pointing and position in the focal plane. Speckle de correlation studies are ongoing [11] and will provide important further understanding of the evolution of aberrations in SPHERE and similar high-contrast systems.

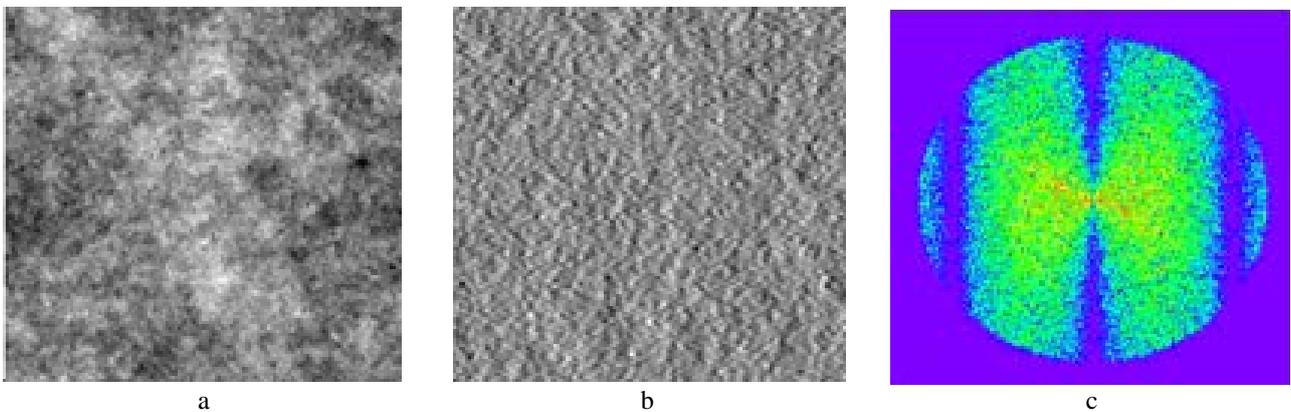
Figure 5. Wave front aberrations appearing due to beam shift. The original phase map (a) with typical $f^{-2}$ PSD is subtracted from itself after a slight horizontal shift giving rise to a flatter wave front dominated by high frequency ripple (b). The PSD of the difference wave front shows the typical butter fly shape which is flat for low frequencies in the horizontal direction.

Thanks to the installation of the Zernike sensor for low-level differential aberrations (ZELDA) [12] in the infrared coronagraph wheel of SPHERE, we have got a good understanding of residual aberrations in the instrument [13]. As seen in Figure 6, the system suffers from poor calibration of low-frequency aberrations, dominated by about 100nm PTV of astigmatism. The reason for this is that the nominal phase diversity-based calibration scheme [14] had to be abandoned due to its incompatibility with the dead actuators of the DM. At the moment of commissioning the presence of these aberrations was found to be acceptable since well tolerated by the apodized Lyot coronagraphs [15] used in the instrument. The ZELDA measurements also show considerable mid-frequency aberrations, however, and a first attempt at calibrating the system using the ZELDA measurements [13] indeed show a significant contrast improvements in this range, see Figure 7.

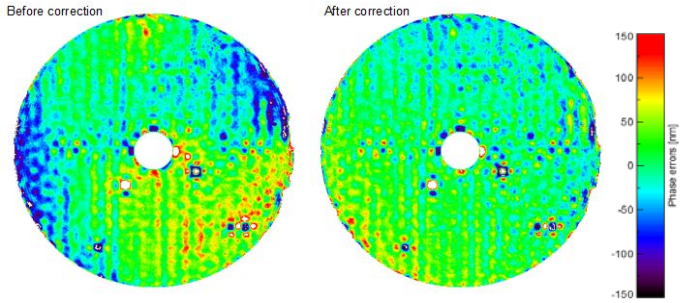 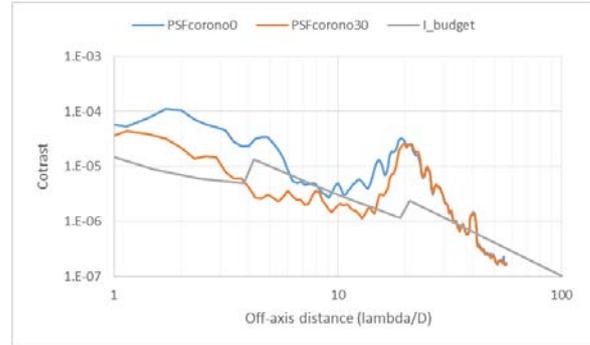

Figure 6. The SPHERE wave front as measured using the ZELDA phase mask wave front sensor located in the infrared coronagraph wheel after running the standard calibration template (left) and after applying a low-pass filtered version of this wave front to the AO reference slopes (right).

Figure 7. Comparison of the expected coronagraphic PSF profile estimated from the wave front budget (grey line) with the actually obtained azimuthally averaged coronagraphic profile using the standard calibration (blue) and the ZELDA calibration (orange).

As seen from these observations, in important improvement of achievable contrast can be reached by implementing NCPA calibrations using the ZELDA sensor. Work is ongoing in collaboration between ESO and the consortium to implement this new feature into the routine operation scheme. One of the main outstanding questions is the lifetime of such calibrations, in particular for the mid-frequency range which will be affected by the rotating ADC prisms in the differential path and the variations in beam shift throughout the common path optics due to atmospheric refraction. The ongoing studies initiated by Milli et al [11] will provide answers to this, but still, it is clear that important improvements in the low spatial frequencies will be gained by implementing ZELDA calibration on a daily basis or even at the beginning of each observation.

A more aggressive approach would be to implement on-line NCPA calibration by replacing the current differential tip-tilt sensor by a ZELDA-based sensor, which, while still allowing precise tip-tilt control, provides real time analysis of the infrared wave front just upstream of the coronagraph [12]. There will still be NCPA residuals between the ZELDA phase mask and the coronagraph mask, put these will be stable and can be calibrated with good precision and long lifetime. We are investigating the opportunity for such a hardware upgrade both in term of feasibility and project structure.

An additional and important advantage of implementing an on-line ZELDA sensor is its capacity to measure accurately the low-wind effect (LWE) [16], offering the possibility to detect and correct in real time the occurrence of this effect which is jeopardizing around 20% of the available observing time.

## 3. TRANSMISSION BUDGET

Faced with the practical difficulties and uncertainties involved in absolute transmission measurements of a complete imaging system, it was agreed to verify the instrumental transmission on a component basis. A complete transmission model of the instrument was built at the start of the project, using typical surface-by-surface transmissivity and reflectivity curves. The budget was updated in the final design phase using quoted values from identified vendors and finally during the manufacturing phase using actual measurements of individual components. This approach allowed early identification of critical components and to successfully fulfil the instrumental specifications. It also allowed for some important system-level trade-offs, such as introducing an acceptable loss of blue transmission of the instrument's entrance window in order to achieve appropriate ghost control across the full waveband of the instrument.

As seen in Figure 8 and Figure 9, the specifications were successfully reached for common path optics as well as for the visible channel ZIMPOL and the near infrared channel IRDIS. Losses due to diffraction and finite fill factor in the micro lens arrays which had not been taken into account in the preliminary studies caused the integral field spectrograph (IFS) to come short of its specified level, but its final performance is still fully acceptable.

Including coronagraphic components and detector quantum efficiencies in the budget allows estimating the complete throughput of the instrument channels, see Figure 10, reaching around 10% in the infrared and almost 30% in the visible. An attempt at comparing this prediction with on-sky observations was made by observing a photometric calibration star during commissioning using the four broad-band filters of the IRDIS camera, see Figure 11. The observed efficiency also includes atmosphere and telescope, and, considering the fact that the telescope primary mirror was in a rather poor state due to a problem with the coating plant and that the atmosphere was not photometric, we consider these measurements to validate our budgeting approach.

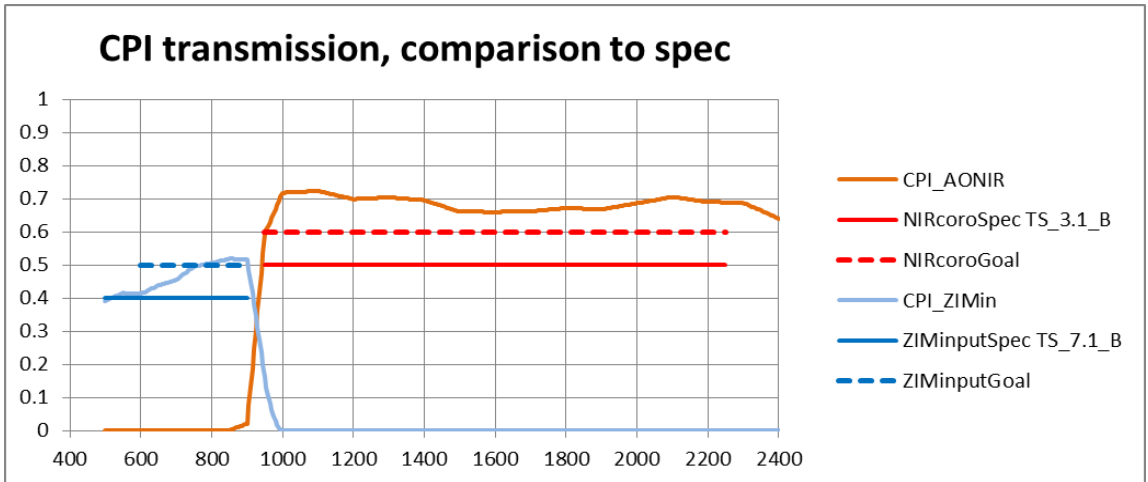

Figure 8. Transmission budget for the common path of the SPHERE instrument from the telescope focus to the interface with the science channels in the visible (light blue line) feeding the ZIMPOL camera and the infrared (orange line) feeding the IRDIS camera and the IFS integral field spectrograph. The dark blue and red lines indicate specified (full lines) and goal (broken lines) transmission levels.

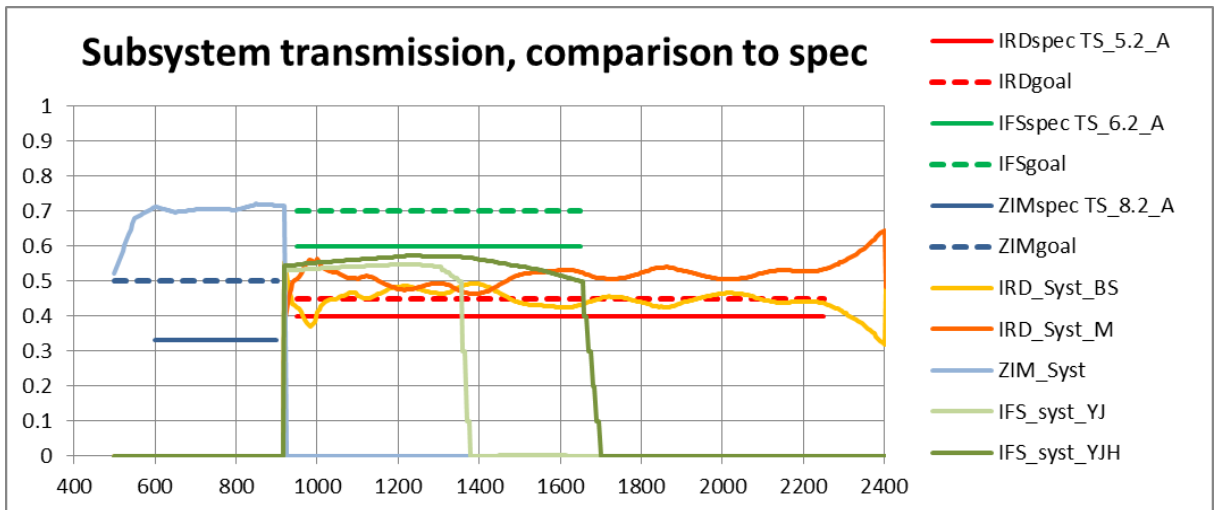

Figure 9. Transmission budget for the SPHERE science channels ZIMPOL (light blue) in the visible and IRDIS (orange and yellow for each of the two dual imaging channels) and the IFS (light and dark green for the YJ and YJH channels, respectively) in the infrared. The dark blue, red and light green lines indicate specified (full lines) and goal (broken lines) transmission levels for each of the channels.

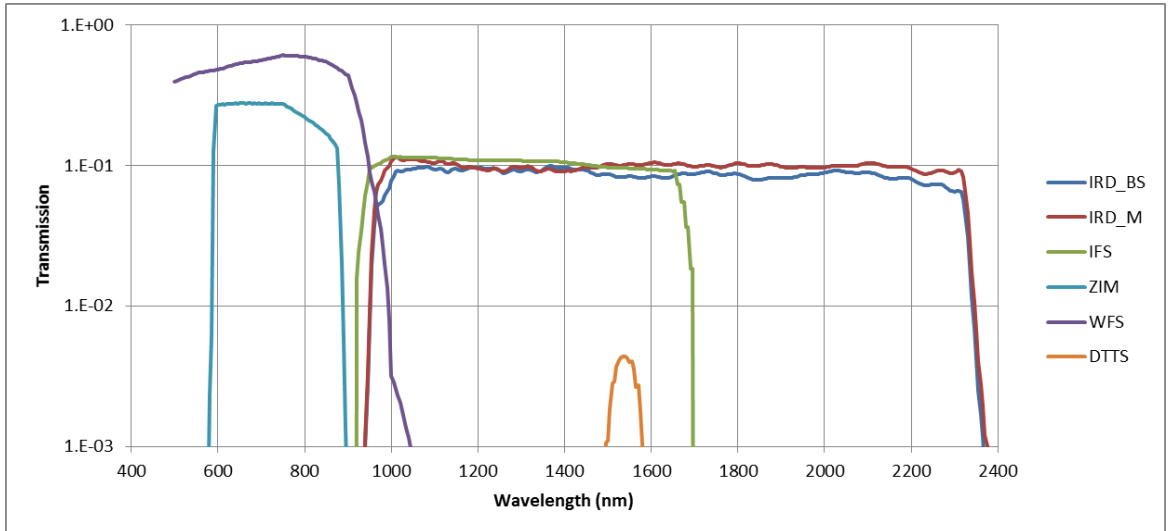

Figure 10. System efficiency up to and including detectors for all the SPHERE channel. The system setup includes typical cornagraphs (classical Lyot corongraph in the visible and an apodized Lyot corongraph in the NIR) and broad-band filters. Fictious, all-transmissive and all-reflective separators between WFS (violet curve) and ZIMPOL (light blue curve), and between IRDIS (red and dark blue curves) and IFS (green curve) are used to illustrate simultaneously all the channels. The orange curve centred around 1550nm represents the efficiency of the differential tip-tilt sensor (DTTS).

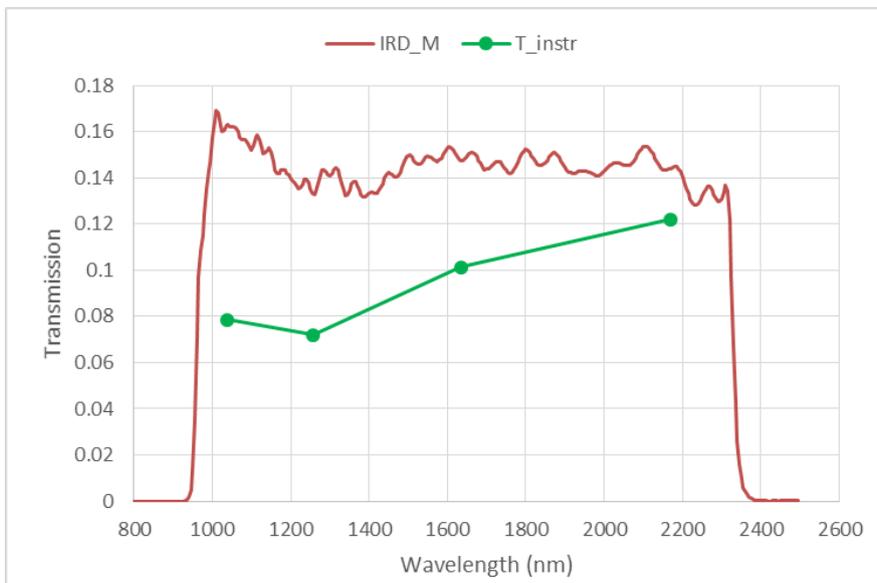

Figure 11. Comparison between instrument transmission (including atmosphere and telescope) estimated from observation of a photometric standard star (green) with as-built budgeted transmission of the IRDIS infrared camera (red).

## 4. DEFORMABLE MIRROR DEFECT MITIGATION

The SPHERE deformable mirror (DM) is a unique and unprecedented piece of opto-electro-mechanics. Developed by the CILAS company in a shared-risk R&D effort led by ESO and financed by the EU, this 180 mm-diameter mirror is based on piezo-stack technology and achieved, at the time of its delivery in 2007, a stunning 10.2 µm MPV (mechanical peak-to-valley) stroke of which less than 15% was required to correct its own deformations. All the 1377 actuators were reported functional.

## 4.1. Dead actuators

We lost several actuators during testing of the system in Europe. The exact mechanism responsible for these losses is not well known, but air-born humidity appears to be at the origin. Also, edge actuators were more prone to dying than others, so although a total of 17 actuators were dead or weakly responding ("zombies") by the time the instrument was shipped to Chile, only 6 of them were seen within the Lyot stop's useful aperture, see Figure 12 (a). The presence of these bright spots in the corongraphic exit pupil created strong fringing in the coronagraphic images (Figure 12 b), and to minimize this effect, new Lyot stops including dead actuator blockers as experimented by the GPI team [17] were made (Figure 12 c) and installed in the instrument during assembly in Paranal. This modificaton was highly successful and no report of fringing has been filed since then.

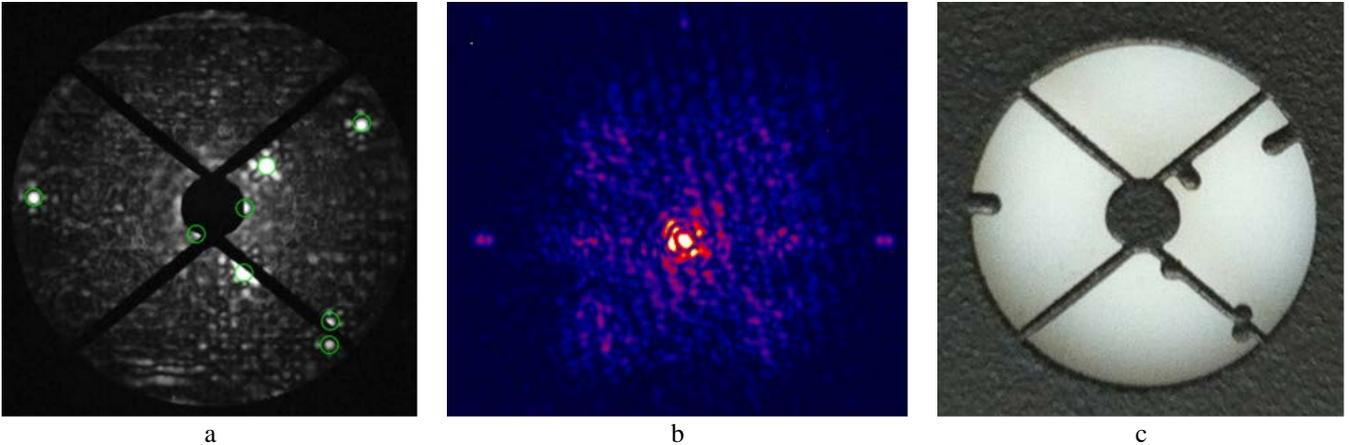

Figure 12. Six dead actuators were seen through the Lyot stop during AIT in Europe (a) causing strong fringe patterns in the coronagraphic image (b). Modified Lyot stops (c) were installed during assembly of the instruments at Paranal.

## 4.2. Astigmatic shape evolution

By the time of its implementation in the SAXO test bed in Paris in 2010, the shape of the DM had evolved, as seen in Figure 13,. A significant amount of astigmatism was then measured on the surface, eating up more than half of the stroke. While this evolution had already started before delivery but was not considered at the time since all measurements were well within the requirements, it was now too large for the mirror to be useful for atmospheric turbulence correction. Furthermore, the astigmatism was found to be variable with temperature, with a sensitivity of about 0.5 μm MPV/K.

Compensatory action using cylindrical lenses was implemented for the test bed, but for a permanent mitigation in the SPHERE system, it was decided to apply cylindrical deformation to one of the three toroidal mirrors polished at LAM by actively stressed substrate polishing [3] thanks to the installation of a deformation harness within its support cell. Used passively during testing in Grenoble, it was motorized before shipping to Chile and an active control algorithm was successfully implemented at the telescope. An iterative fine adjustment procedure was preceded by an initial adjustment using a thermal control law of form

$$TM3enc = (T - T0)\ C,$$

where T is temperature, T0 is the zero encoder temperature and C is the thermal sensitivity.

But the DM of SPHERE had not said its last word. Only a few weeks after first light, during the second commissioning run in July 2014, it was found necessary to modify T0 from its initial value of 5.7°C to 7.5°C, and during the fourth commissioning run in September it was again increased to 11°C. Under the rejuvenating action of the dry Paranal atmosphere the DM was flattening towards its original shape, as seen in Figure 13, leading to a rapid evolution of the T0 value (Figure 16). As a consequence, since the TM3 deformation harness was only built to provide positive ("push") deformation, it was predicted that at the onset of the Chilean winter of 2015 it would be impossible to compensate properly the DM astigmatism for temperatures below 13°C.

For historical reasons, a spare TM3 had been polished, and it was decided to equip it with a modified harness allowing both "push" and "pull" deformation [18]. Using the same motorization and safety features as the old harness, this new system was built to be perfectly compatible with the old system in terms of optical, mechanical, electrical, and control interfaces, only requiring recalibration of the control law parameters. This new system was built during the first months of 2015 and the replacement took place successfully in May 2015, see Figure 15 [18].

Although the evolution of the DM has slowed down, it is still ongoing. A final adjustment of T0 was made in January 2016, setting it to its current value of 14°C.

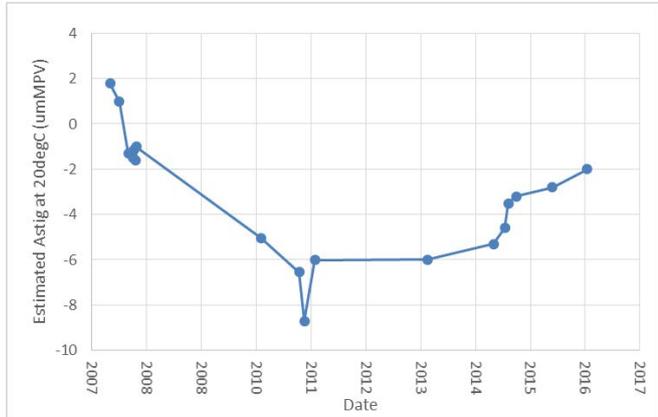

Figure 13. Long-term evolution of DM astigmatism. The values reported from 2007 are deduced from the manufacturer's measurement report. Values from 2010 and 2011 were made during tests at the SAXO test bench in Paris; the outlier value from December 2010 was made during thermal testing. The value reported from 2013 was deduced from tests during validation in Grenoble, and the remaining points are from installation and operation at Paranal.

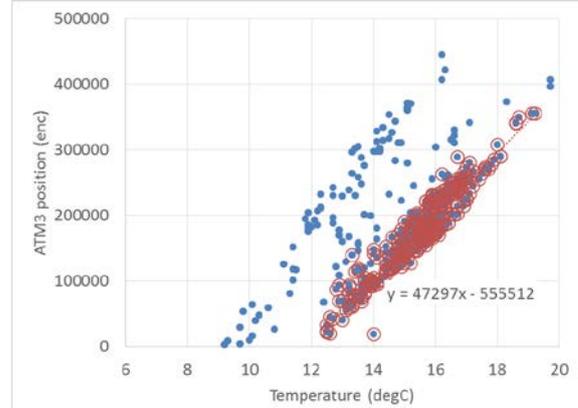

Figure 14. Evolution of the zero encoder temperature of the TM3 harness compensating the DM thermal sensitivity during the first year of operation. Blue dots plot the TM3 encoder position as a function of temperature after daily calibrations. The red circles correspond to calibrations done after 1/11/2015.

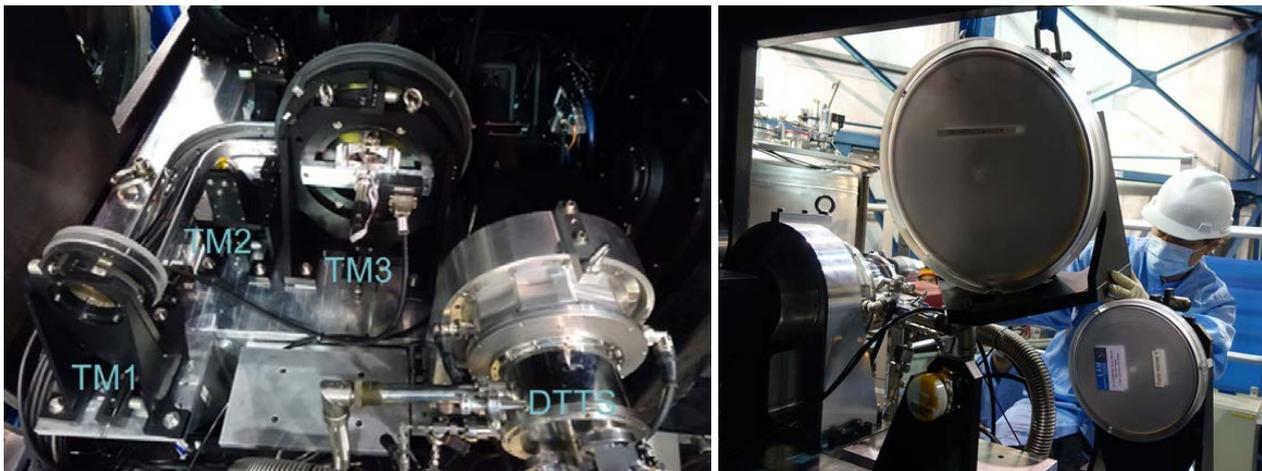

Figure 15. View of the interior of SPHERE showing the three toroidal mirrors, TM1, TM2, TM3, and the differential tip-tilt sensor (DTTS) (left) and the TM3 seen during its craning out of the instrument for replacement in May 2015 (right).

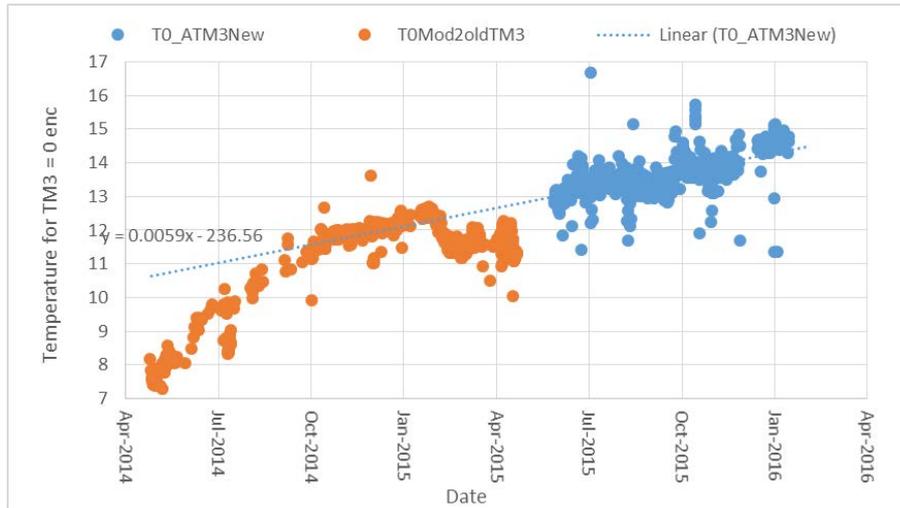

Figure 16. Evolution of the T0 value, corresponding to the temperature for which the TM3 motor would need to be set to zero encoders for optima compensation of the DM astigmatism for the old, push only TM3 harness (orange) and for the new, push and pull harness (blue). The temporary reduction in T0 around March 2015 coincides with a high-humidity period on the summit.

## 5. CONCLUSION

Careful system design and state-of-the-art manufacturing techniques have led to the successful implementation and high performance of the SPHERE which was installed on the Nasmyth platform of VLT UT3 in May 2014. We have described two of the main performance budgets, transmission and wave front, and compared them with on-sky measurements, finding good correspondence in both cases. We have seen that significant improvements can be made by implementing NCPA calibration using the ZELDA Zernike phase mask sensor already present in the instrument. In spite of its imperfections, the SPHERE deformable mirror performs as required thanks to important mitigation efforts. Lyot stops were replaced during transfer to Paranal, and the motorization of the large toroidal mirror allows efficient compensation of the thermally and temporally varying DM shape, hence recovering the full dynamic range of the mirror. We have retraced the history of these shape variations and described the steps of the compensation implementation.

## ACKNOWLEDGEMENTS

SPHERE is an instrument designed and built by a consortium consisting of IPAG, MPIA, LAM, LESIA, Laboratoire Fizeau, INAF, Observatoire de Genève, ETH, NOVA, ONERA and ASTRON in collaboration with ESO. We are grateful to all the consortium members who contributed to build, test, and validate the instrument in Europe and in Chile. We also thank the ESO staff involved with the installation and operation of SPHERE on the VLT-UT3.